\documentclass[runningheads]{llncs}

\usepackage{graphicx}
\usepackage{listings}
\usepackage{varwidth}
\usepackage{multirow}
\usepackage{scalerel}
\usepackage{amsmath,amssymb,amsfonts}
\usepackage{algorithm} 
\usepackage{graphicx}
\usepackage{textcomp}
\usepackage{listings}
\usepackage{graphicx}
\usepackage{float}
\usepackage{algpseudocode}
\usepackage{rotating}
\usepackage{amsmath}
\usepackage{verbatim}
\usepackage{framed}
\usepackage{balance}
\usepackage{tabu}
\usepackage{url}
\usepackage{tikz}

\usetikzlibrary{svg.path}
\definecolor{orcidlogocol}{HTML}{A6CE39}
\tikzset{
  orcidlogo/.pic={
    \fill[orcidlogocol] svg{M256,128c0,70.7-57.3,128-128,128C57.3,256,0,198.7,0,128C0,57.3,57.3,0,128,0C198.7,0,256,57.3,256,128z};
    \fill[white] svg{M86.3,186.2H70.9V79.1h15.4v48.4V186.2z}
                 svg{M108.9,79.1h41.6c39.6,0,57,28.3,57,53.6c0,27.5-21.5,53.6-56.8,53.6h-41.8V79.1z M124.3,172.4h24.5c34.9,0,42.9-26.5,42.9-39.7c0-21.5-13.7-39.7-43.7-39.7h-23.7V172.4z}
                 svg{M88.7,56.8c0,5.5-4.5,10.1-10.1,10.1c-5.6,0-10.1-4.6-10.1-10.1c0-5.6,4.5-10.1,10.1-10.1C84.2,46.7,88.7,51.3,88.7,56.8z};
  }
}

\newcommand\orcidicon[1]{\href{https://orcid.org/#1}{\mbox{\scalerel*{
\begin{tikzpicture}[yscale=-1,transform shape]
\pic{orcidlogo};
\end{tikzpicture}
}{|}}}}
 
\usepackage{todonotes}
\newboolean{showcomments}
\setboolean{showcomments}{true} \ifthenelse{\boolean{showcomments}}
{\newcommand{\nb}[2]{
  \fcolorbox{black}{yellow}{\bfseries\sffamily\scriptsize#1}
  {\sf\small$\blacktriangleright$\textit{#2}$\blacktriangleleft$}
 }
 
}
{\newcommand{\nb}[2]{}
 
}

\usepackage{listings}
\usepackage{subfig}
\usepackage{graphicx}
\usepackage{xcolor}
\usepackage{multirow}
\usepackage{setspace}
\usepackage{amssymb}
\usepackage{array}

\lstdefinelanguage{json}{
	basicstyle=\footnotesize\ttfamily,
	numbers=left,
	numberstyle=\scriptsize,
	stepnumber=1,
	numbersep=8pt,
	showstringspaces=false,
	breaklines=true,
	frame=lines,
	backgroundcolor=\color{background},
	literate=
	*{0}{{{\color{numb}0}}}{1}
	{1}{{{\color{numb}1}}}{1}
	{2}{{{\color{numb}2}}}{1}
	{3}{{{\color{numb}3}}}{1}
	{4}{{{\color{numb}4}}}{1}
	{5}{{{\color{numb}5}}}{1}
	{6}{{{\color{numb}6}}}{1}
	{7}{{{\color{numb}7}}}{1}
	{8}{{{\color{numb}8}}}{1}
	{9}{{{\color{numb}9}}}{1}
	{:}{{{\color{punct}{:}}}}{1}
	{,}{{{\color{punct}{,}}}}{1}
	{\{}{{{\color{delim}{\{}}}}{1}
	{\}}{{{\color{delim}{\}}}}}{1}
	{[}{{{\color{delim}{[}}}}{1}
	{]}{{{\color{delim}{]}}}}{1},
}
\usepackage{hyperref} 

\begin{document}

\title{Blockchain-Oriented Services Computing in Action: Insights from a User Study}	

\author{
Giovanni Quattrocchi
\inst{1}
\and Damian Andrew Tamburri
\inst{2,3}
\and \\ Willem-Jan Van Den Heuvel \inst{3,4}
}
\authorrunning{G. Quattrocchi, D. Tamburri, W. Van Den Heuvel }
\institute{Politecnico di Milano, Italy \\
\and 
Eindhoven University of Technology, Netherlands \and
Jheronimus Academy of Data Science, Netherlands \and
Tilburg University, Netherlands \\
\email{giovanni.quattrocchi@polimi.it}\\
\email{d.a.tamburri@tue.nl}\\
\email{W.J.A.M.v.d.Heuvel@jads.nl}\\
}

\titlerunning{Blockchain-Oriented Services Computing in Action}

\maketitle

\begin{abstract}
Blockchain architectures promise disruptive innovation but factually they pose many architectural restrictions to classical service-based applications and show considerable design, implementation, and operations overhead. Furthermore, the relation between such overheads and user benefits is not clear yet. To shed light on the aforementioned relations, a service-based blockchain architecture was designed and deployed as part of a field study in real-life experimentation. An observational approach was then performed to elaborate on the \emph{technology-acceptance} of the service-based blockchain architecture in question. Evidence shows that the resulting architecture is, in principle, not different than other less complex equivalents; furthermore, the architectural limitations posed by the blockchain-oriented design demand a significant additional effort to be put onto even the simplest of functionalities. We conclude that further research shall be invested in clarifying further the design principles we learned as part of this study as well as any trade-offs posed by blockchain-oriented service design and operation. 
\end{abstract}

\section{Introduction}

Blockchain technology is heralded as a silver bullet for a wide range of problems, yet the stylistic restrictions posed on top of more classical service-oriented architectures \cite{richards2016microservices} that blockchain-oriented service design forces into the equation limit the throughput and latency of blockchain transactions \cite{swan2015blockchain}. For example, the Bitcoin network \cite{LischkeF16} can currently handle a maximum of 7 transactions per second, although the Ethereum network \cite{WesslingG18} offers a relatively higher number of 15 transactions per second (tps); to date, this rate is not compatible to the processing capacity of other networks such as VISA (2,000tps) and Twitter (5,000tps). Creating a new block that is required to assure the safety of the network requires 10 minutes which significantly slows down the time to complete one transaction, resulting in low latency. 

As such, the usability of blockchain designs may need further attention than software designs following other architecture patterns \cite{BMRSS96}. More specifically, we are interested in studying the extent to which the intrinsic limitations of blockchain-oriented designs weigh on their perceived end-user effectiveness \cite{yli2016current}. To look into blockchain usability from a design perspective, this article offers an empirical \emph{invivo} field study designed using the guidelines defined in the Technology Acceptance Model (TAM) \cite{Bradley2009} and related frameworks \cite{2collab178971}. First, we designed and prototyped a blockchain-based service-oriented transactional architecture. Second, we deployed and tested the architecture from the user perspective, by interviewing its end-users as part of a controlled experiment.

The results indicate that end-users do perceive several advantages (i.e., good information and transactions transparency, ease-of-use as well as user-friendliness) from using the blockchain but also that the blockchain imposes a \emph{lock-in} which even comes at a cost of +30\% development times and efforts. While the alternative transactive methods are perceived as no longer usable or obsolete by users, service architects may have to evaluate the resulting trade-offs a bit more carefully. Overall, these findings indicate a need to further understand the process of designing for blockchain-oriented service engineering \cite{WesslingG18}.

The practical implications are at least twofold: on one hand, blockchain limitations need to be overcome with technical and design devices capable of addressing them, on the other hand, the risks of not involving end-users in the design might lead to the undesired circumstances we report in the results section, e.g., the lock-in condition we reported from our user study.

The rest of this paper is structured as follows. First, Sec. \ref{rw} outlines the related work. Beyond that, Sec. \ref{rd} outlines our research design, also providing a birds-eye view of the architecture we designed and implemented as a field-study device. Further on, Sec. \ref{eval} provides the \emph{enfield} deployment and experimentation over our blockchain-oriented prototype. Finally, Section \ref{conc} concludes the paper. 
\section{Related Work}
\label{rw}

Technology acceptance by end-users is a well-established concept, and has been studied extensively in prior research \cite{venkatesh2003user}. One of the most used lenses to study technology acceptance is the well-known Technology Acceptance Model (TAM) \cite{Bradley2009}. Results from studies that employed the TAM suggest that when users are presented with new technology, at least two factors influence their decision about how and when they use it, namely: (1) perceived usefulness (PU) --- defined by Davis \cite{Davis89} as ``the degree to which a person believes that using a particular system would enhance his or her job performance"; (2) perceived ease-of-use (PEOU) --- defined by Davis \cite{Davis89} as ``the degree to which a person believes that using a particular system would be free from effort". In their variant DeLone and McLean \cite{delone2003delone} introduce two additional variables: (3) perceived information quality (PIQ) and (4) user satisfaction. Although technology acceptance of the end-users of various IS has been thoroughly studied, except for a study conducted by Folkinshteyn and Lenon \cite{folkinshteyn2016braving}, no research has been conducted to gauge users' acceptance of blockchain technology. Moreover, the study carried out by Folkinshteyn and Lenon is limited to the use of the Bitcoin protocol and is based on literature rather than capturing the perceived user perception in practice.

\section{Research Design}
\label{rd}
The problem addressed in this paper reflects the shortage of information concerning the user-acceptance of blockchain architectures. On one hand, such architecture poses a considerable strain on designers given their constraints and architectural limitations \cite{WesslingG18}. On the other hand, the end-user benefits and, more specifically, the \emph{technology acceptance} \cite{Kakar17} from such end-users is questionable at best. With technology acceptance, we indicate the information systems' architecture approach that focuses on establishing how users come to accept and use a specific technological architecture.

Given its early stage of adoption, we aim to articulate the effect of blockchain limitations to the above-mentioned dimensions of technology acceptance. More specifically, we pose the research questions below:

\begin{enumerate}
\item[RQ1] \emph{To what extent do blockchains enhance service application usefulness?}
\item[RQ2] \emph{To what extent is the blockchain transparent to direct service use?}
\end{enumerate}

With the above RQs, we aim at understanding the \emph{usefulness} and \emph{transparency} of the blockchain design principles and restrictions; in so doing, we prototype a blockchain transaction system and execute a field-study featuring \emph{enfield} questionnaires and web-surveys designed to evaluate usefulness as an essential dimension of the study.

\subsection{Blockchain Technology Acceptance: Field Study Design}
\label{experiment}

As previously mentioned, to attain our research results we conducted a field study using a \emph{enfield}-deployed version of our research prototype. The prototype in question was deployed as production-ready --- meaning that the prototype was in fully-working conditions and has been deployed in \emph{practice} on several similar occasions (e.g. other festivals). The field experiment followed the guidelines of Singer et al. \cite{LES05} and essentially involved: (1) end-use of the system in the context of a real-life event involving the active use of the prototype; (2) follow-up interviews featuring a web survey which followed a random sampling approach. 

The data used for this research was collected during the pre-edition of a festival in the Netherlands\footnote{For more information about the festival the reader should visit: \url{https://www.welcometothevillage.nl}}. The festival annually hosts 3000+ visitors and 200+ volunteers who co-organize the festival. The volunteers were all asked to participate in this research in an \emph{opt-in} fashion; the involvement in the context of this study was featuring the use of our blockchain platform. The platform in question allowed the participants of this research to (a) buy beverages using a token that was created on the blockchain platform and converted from real cash; (b) buy tokens with euros directly on the platform and upload them to their account using a QR-code wristband; (c) finally, check their account balance. Users can download a mobile app or surf an internet page to access the aforementioned features.

The study involved a total of 48 randomly-sampled end-users with a mean age of 23 (standard deviation of 4,37). The population involved a total of 23 male subjects and 18 female subjects. The population is skewed towards more tech-savvy people with a ratio of ~1:3. The respondents for this study were administered a questionnaire consisting of 11 items to measure certain aspects detailed in the following. All questions were addressed by the studied subjects along a typical 5-factor Likert-scale \cite{GilG12} allowing for subsequent content analysis \cite{content}. For the sake of reference and replicability, the questionnaire is available online\footnote{\url{https://tinyurl.com/ycha8282}}. All of the respondents filled in the survey albeit not completely.

The questions used for the survey (after the use of the platform was recorded) were derived from \cite{rai2002assessing}, which were in turn based on the DeLone and McLean model. Users Acceptance Testing questions are designed to evaluate the field use of the proposed technology along the typical criteria defined for technology acceptance, as defined previously in the prior sections. 

\begin{enumerate}
\item \textit{Perceived information quality} provided by the system (4 items), defined as the extent to which the system provides the respondents with accurate information, delivered in the format required by the users about their account balance,
\item \textit{Perceived ease-of-use} as a concept to measure the systems' quality (1 item), which can be referred to as the degree to which the SIS is easy to use when making transactions to buy beverages and, 
\item \textit{Perceived usefulness of the system}, to gauge the perceived benefit of using the system (4 items), which we define for this research as the degree to which the user believes that the system caters to them in buying beverages, along with,
\item \textit{Perceived user satisfaction} with regards to the architecture quality measured by one item evaluating the users' technical feedback on the architecture features \cite{Dhungana06} they tried out during their experimentation.

\end{enumerate}

To rule out rivaling explanations for the results of the study, we have included two control variables in our model: 

\begin{enumerate}
    \item \textit{Utilization}, defined as the degree to which a user is dependent on the IS to carry out his or her tasks, has been included to rule out that the users' perception of the system can be attributed to the fact that they cannot work around the system. One item was included to measure this variable
    
    \item \textit{Technology Savviness} that is, questions designed to evaluate the users' perceived confidence with software technology along 5 evaluation criteria, namely, Search Engines (confidence with information retrieval and storage), Social Media (confidence with social networks, and online digital presence), Digital Content (confidence with knowledge-bases as well as general information management with different document formats), Software Security (knowledge and confidence against anti-viral, malware or other online security threats) and Software Care (confidence with software repair);
\end{enumerate}

\begin{figure}[t]
	\centering
	{\includegraphics[width=0.75\columnwidth]{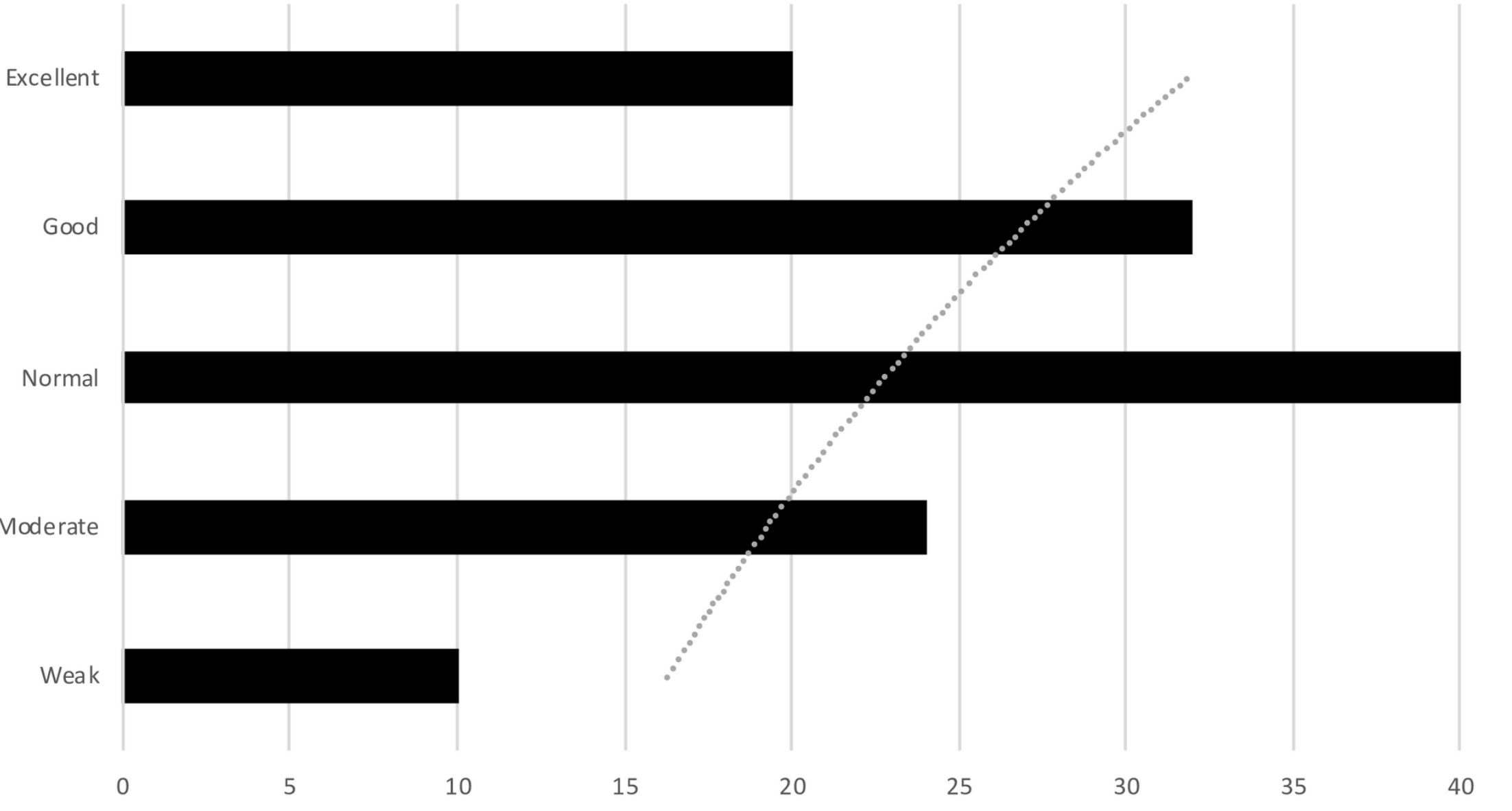}}
	\caption{Architecture information and transactions quality, that is, the extent to which the blockchain-oriented design is perceived as providing appropriate and useful transactions information; the y-axis indicates likert-scale levels while x-axis indicates \#respondents.}
	\label{rq_1}
\end{figure}

The data we obtained in our end-user survey featured 48 timestamped responses along a 5-dimensional Likert scale. To analyze available data, regular statistical modeling was adopted along with non-boosted logistic regression modeling. More specifically, in the scope of RQ1, namely, \emph{to what extent do blockchains enhance information system usefulness?}, we produced combined Likert-scale responses to three specific questions. On one hand, we summed results from two questions: (1) \emph{``the blockchain architecture offers me all the information I require to perform my transaction"}; (2) \emph{``the possibilities that the blockchain platform offers to increase my balance are satisfactory"}. Subsequently, to avoid observer bias \cite{simonssources2009}, we triangulated the two questions above with a control question, namely, \emph{``the possibilities that the blockchain architecture offers to perform transactions are satisfactory"}; thus, the sum obtained above was decreased with the Likert-scale results from the above  question. The results were plotted using a bar chart and a logistic regression trendline was fitted with the data (see Fig. \ref{rq_1}).

Finally, in the scope of RQ2, namely, \emph{to what extent is the blockchain transparent to direct use?}, we computed Pearson's product-moment correlation coefficient between the Likert-scale responses for question \emph{``making transactions using the blockchain platform is easy and transparent."} and question \emph{``I am dependent on the blockchain platform to perform my transactions."} --- our research assumption is that the significant correlation between the two responses indicates a strong dependency on the blockchain platform to perform transactions with respect to the regular transactional alternative.

\section{Research Results}
\label{eval}

Figure \ref{rq_1} shows our results in the context of RQ1. More specifically, the figure shows a linear trend with respect to responses concerning architectural information and transaction quality within the blockchain-oriented design under study. Although the trend pends slightly by about 3\% towards a positive transactional information quality, the data shows a rather inconclusive outcome with respect to the extent that the presence of a blockchain-oriented design reinforces transactions' quality. Our data indicate that the trend seems positive but there is little to no indication that the trend is connected to the blockchain and the effect size we report is non-significant.

\begin{figure}[t]
	\centering
	{\includegraphics[width=0.7\columnwidth]{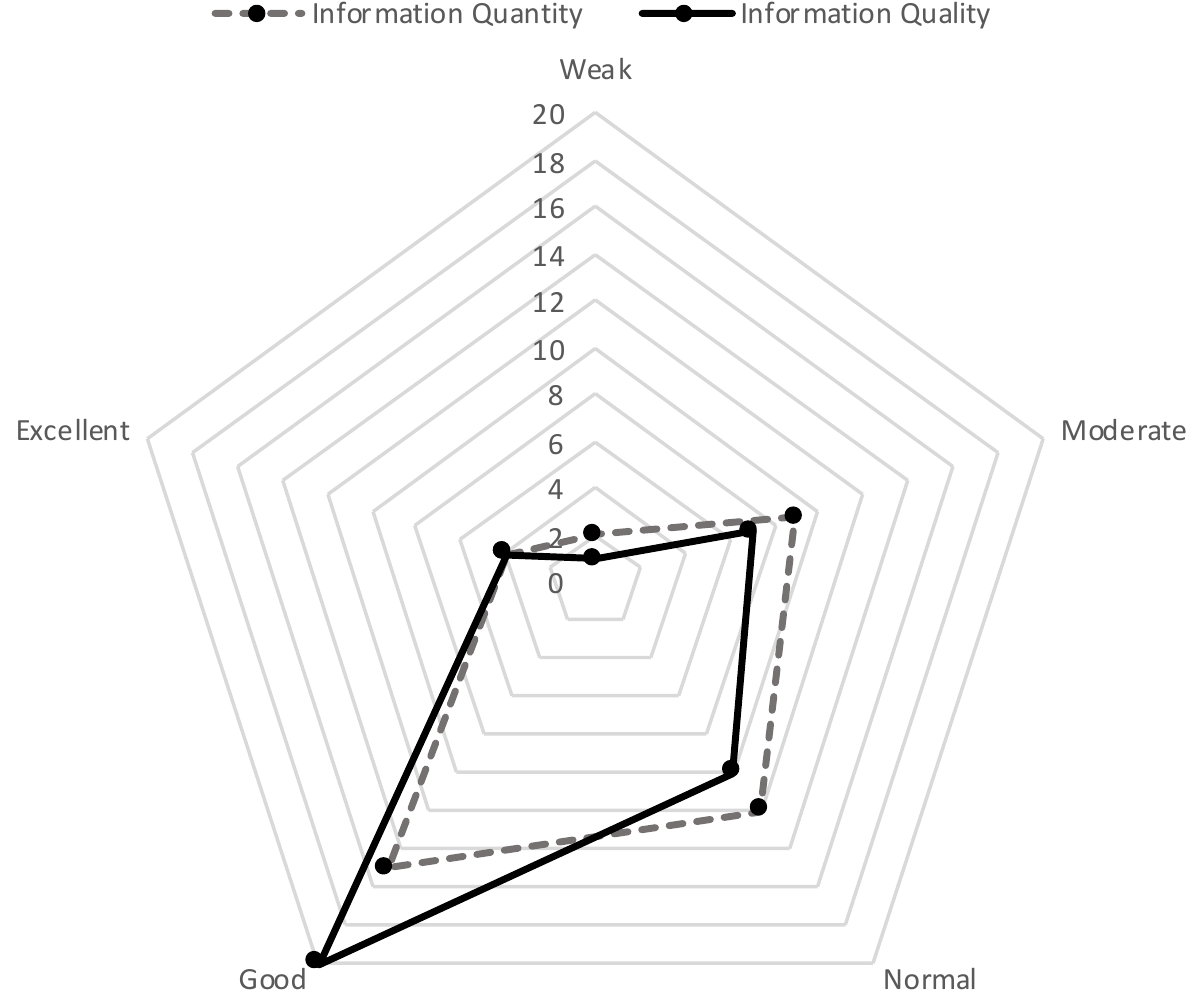}}
	\caption{Information Quantity vs. Quality, that is, the extent to which the blockchain-oriented design yields an appropriate quantity of information with respect to an appropriate quality of presented information --- the data highlights a 78\% overlap.}
	\label{rq1_ab}
\end{figure}

With respect to transactions information quality and quantity in the scope of the blockchain-oriented design under study, Fig. \ref{rq1_ab} depicts a radial diagram to capture the overlap in question. The figure shows a definitive overlap between information quantity (dotted, smaller inner-line on the figure) and information quality (continuous black line). The overlap rests around 78\% indicating a considerable perceived overlap between quality and quantity of information in the context of performed transactions by the users in our study --- this extent of overlap between information quantity and quantity suggests that the presence of the blockchain underneath the implementation under study makes information and transactional quality/usefulness more explicit.

\begin{figure*}[t]
	\centering
	{\includegraphics[width=0.80\columnwidth]{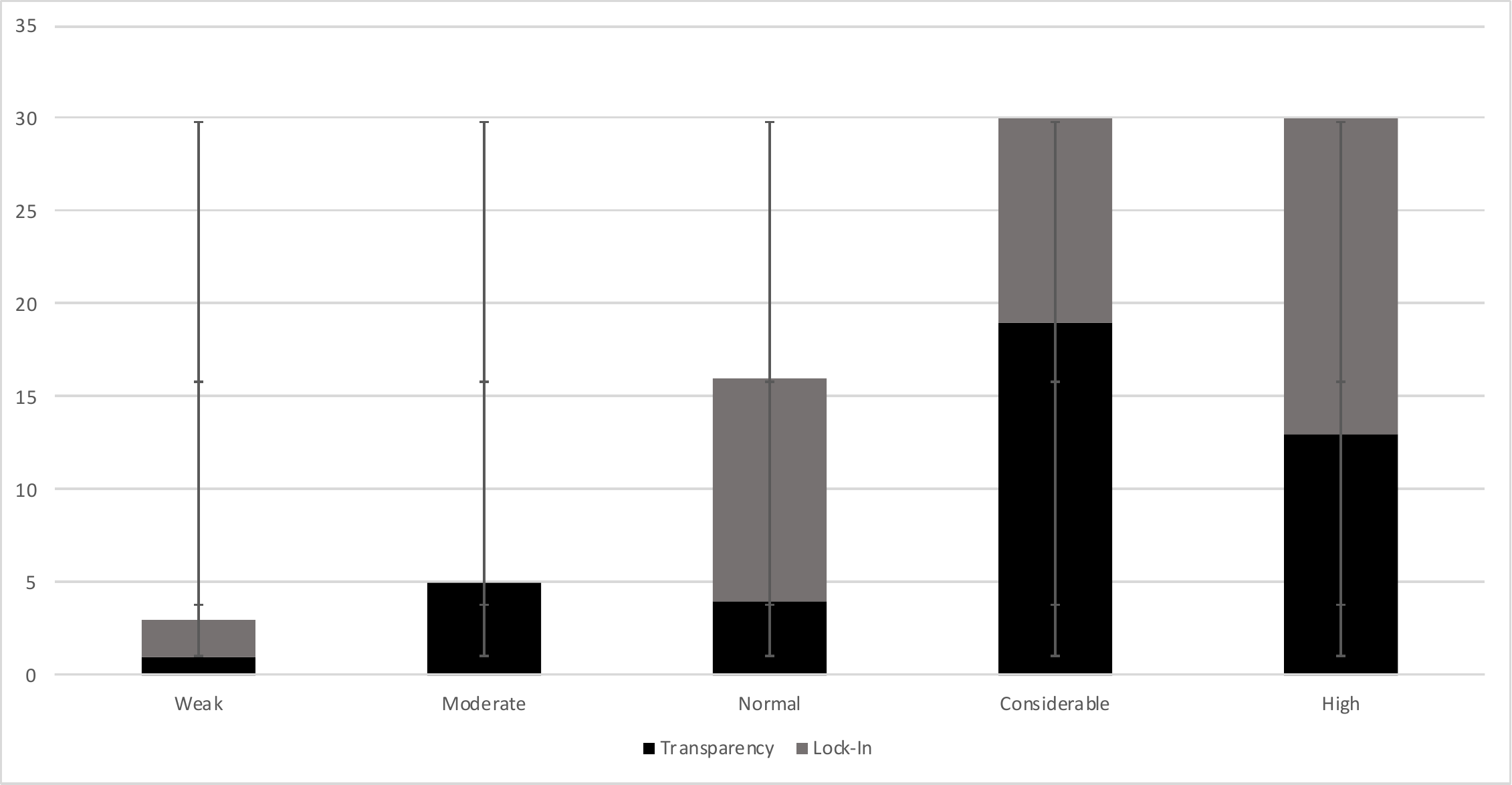}}
	\caption{perceived information transparency --- the plot remarks the negative trend in transparency (black-area in the box-plot) with respect to perceived lock-in (greayed-out area in the box-plot).}
	\label{rq2}
\end{figure*}

In terms of transparency, the data shows a mild correlation of ~0.39 (P-value 0.012312$<<$0.05) between the extent to which the transactions performed by means of the proposed blockchain-oriented design are transparent and the extent to which the user feels constrained to use the blockchain-oriented design \emph{only}, that is, instead of the classical transaction alternative. Concerning this apparent \emph{lock-in} phenomenon, Fig. \ref{rq2} provides an overview of the compulsive mutual effect size in the scope of the aforementioned lock-in.

The figure shows an increasing lock-in perception with a plateau around the \emph{``considerable"} value but with a consequent negative trend immediately afterward with a trade-off drop of 33\%. This indicates that users perceive lock-in over transparency.

\section{Conclusions}
\label{conc}
Blockchain-oriented applications are increasingly picking up interest in the more general audience both from the perspective of practice and academic research inquiry. Little is known, however, over the extent to which blockchain-oriented designs are perceived by their users.

Our study reported that i) blockchain-oriented designs support systems usefulness, and ii) the formation of a significant \emph{lock-in} phenomenon wherefore blockchain users seem to perceive transparency to the extent to which the blockchain becomes a lock-in with respect to conventional transactions system.

\bibliographystyle{plain}

\bibliography{blockchain}

\end{document}